\documentstyle[referee,psfig]{l-aa}

\def\ion#1#2{#1$\;${\small\rm{#2}}\relax}
\def\ionT#1#2{#1$\;${\rm{#2}}\relax}

\def\ltsima{$\; \buildrel < \over \sim \;$}
\def\simlt{\lower.5ex\hbox{\ltsima}}
\def\gtsima{$\; \buildrel > \over \sim \;$}
\def\simgt{\lower.5ex\hbox{\gtsima}}
\def\kms{{\rm\,km\,s^{-1}}}
\def\kpc{{\rm\,kpc}}

\def\msun{{\rm\,M_\odot}}

\def\pc{{\rm\,pc}}

\def\s{\ifmmode \widetilde \else \~\fi}
\def\={\overline}

\def\spose#1{\hbox to 0pt{#1\hss}}

\def\etal{{\it et al.\ }}

\def\eg{{e.g.\ }}
\def\ie{{i.e.\ }}
\def\lta{\mathrel{\spose{\lower 3pt\hbox{$\mathchar"218$}}
     \raise 2.0pt\hbox{$\mathchar"13C$}}}
\def\gta{\mathrel{\spose{\lower 3pt\hbox{$\mathchar"218$}}
     \raise 2.0pt\hbox{$\mathchar"13E$}}}
\def\Dt{\spose{\raise 1.5ex\hbox{\hskip3pt$\mathchar"201$}}}	
\def\dt{\spose{\raise 1.0ex\hbox{\hskip2pt$\mathchar"201$}}}	

\def\=={\equiv}

\def\dotsfill{\leaders\hbox to 1em{\hss.\hss}\hfill}

\def\sgr{the Sagittarius dwarf spheroidal}
\def\sgg{the Sagittarius dwarf}
\def\Sgg{The Sagittarius dwarf}
\def\com{center of mass}

\def\Gyr{{\rm\,Gyr}}

\def\araa{ARA\&A~}		
\def\aj{AJ~}			
\def\aap{A\&A~}			
\def\aaps{A\&AS~}		
\def\apj{ApJ~}			
\def\apjl{ApJ~}			
\def\mnras{MNRAS~}		
\def\nat{Nature~}

\def\pasj{Publ. Astron. Soc. of Japan~}

\newcommand\comment[1]{\null}

\begin{document}

   \thesaurus{10         
              (09.07.1;  
               10.05.1;  
               10.11.1;  
               11.09.2;  
               11.12.1)} 
   \title{Archer of the  Galactic disk? \\ The effect on the outer \ionT{H}{I}
disk of  the Milky Way of  collisional encounters with the Sagittarius dwarf
galaxy}

   \subtitle{}

   \author{R. A. Ibata\thanks{\emph{Present address:} European Southern
Observatory, Karl-Schwarzschild Stra\ss e 2, D--85748 Garching bei M\"unchen, Germany}
\and
           A. O. Razoumov}

   \offprints{R. A. Ibata}

   \institute{
Department of Physics \& Astronomy, University of British Columbia, \\
2219 Main Mall, Vancouver, B.C., V6T 1Z4, Canada \\
Electronic mail: surname@astro.ubc.ca
             }

   \date{Received September 15, 1989; accepted March 16, 1990}

   \maketitle
\markboth{R. A. Ibata, A. O. Razoumov: Outer \ion{H}{I} disk}{}

\begin{abstract}
Hydrodynamical  calculations   undertaken   to  simulate   the   collisional
interaction  between \sgg\ and   the   Galactic outer \ion{H}{I}  disk   are
presented, constrained  by recently derived orbital  and mass parameters for
this  dwarf  galaxy.  It  is found  that   a significant  distortion to  the
structure of  the Galactic \ion{H}{I} disk  will be induced by the collision
if the mass   of  the dwarf exceeds  $   \sim  10^9 \msun$;  this   value is
consistent with an  estimate derived by  requiring that the  dwarf galaxy is
sufficiently   robust to survive   tidal     disruption until the    present
time. Though the precise  details of the interaction  are compromised in our
simulations by the   lack of a live  Galactic  halo, we find  that for model
masses $\simgt 5 \times 10^9 \msun$, prominent spiral arms and a substantial
lopsidedness in  the outer  disk  are produced.   Furthermore,  a noticeable
warp-like   structure   is induced  in   the   disk.   Thus \sgg\  may  have
significantly affected the star formation history and structure of the outer
Galaxy.   These  simulations  confirm  the possibility  of   determining the
current  merging rate of low  surface brightness, gas-poor dwarf galaxies of
mass $\simgt 10^9 \msun$ onto giant spiral galaxies from careful analysis of
observations of the structure of \ion{H}{I} disks.

      \keywords{Galactic dynamics --
                interstellar medium --
                Sagittarius dwarf galaxy
               }

\end{abstract}

\section{Introduction}

The observed warps  of spiral galaxies are   commonly believed to be  either
caused by the torque due to  a misalignment of the outer  halo and the disk,
or to  be  the natural  oscillatory response   of  the  gas  layer to  small
perturbations  (Binney  1992).   However, recent numerical  simulations have
shown that, in models with realistic disk to  halo mass ratios, disks placed
inside oblate halos align themselves with the  plane of symmetry of the halo
within  a  few  orbital periods  (Dubinski \&   Kuijken  1995).   So in both
scenarios  it seems likely   that perturbers are   required to  maintain the
warps.

The influence of specific perturbers on  the disk of the  Milky Way has been
considered. Weinberg (1995)  put forward a model where  disk modes are being
excited  by  the  joint torque  from the   Magellanic Clouds   and the halo.
Another possible perturber  of the  Galactic  disk is the Sagittarius  dwarf
galaxy (Ibata \etal   1994; Ibata  1994), which  likely   collides with  the
Galactic disk every $\sim 1 \Gyr$ (Ibata \etal 1997, hereafter IWGIS).  This
intriguing possibility was brought to attention by  Lin (1996), who realized
that the position   where  one  of the   likely orbits   (given the  initial
kinematic data of Ibata \etal 1994) last passed through the Galactic disk is
coincident   with  the position  where   the  maximum  displacement of   the
\ion{H}{I} warp would have been at that time.

Lin modeled the Galactic \ion{H}{I} gas with 4000 massless and collisionless
tracer particles  placed in circular orbits  in a fixed  Galactic potential.
The response of    these  particles to  the passage   of  a  (collisionless)
point-mass particle  on the orbit deduced for  \sgg\ was then  analyzed.  He
found that, if \sgg\ has a mass of at least $5 \times 10^9 \msun$, and moves
on an  orbit which  fit  best  the   then extant  data, then  the  resulting
perturbation  to the modeled  disk, when  evolved  up  to the  present time,
looked similar  to the  observed warped distribution   of \ion{H}{I} in  the
outer Galactic disk.

However,  by   choosing massless   particles   as tracers  of   the Galactic
\ion{H}{I} disk,  the Lin  (1996) study  did not model   the nature  of that
Galactic  component, which of course  is  both self-gravitating and gaseous.
Furthermore, an analysis  of recent   data (IWGIS)  supports orbits with   a
longer period than that adopted by Lin (1996); the currently best-fit orbit,
implies that \sgg\ did not collide with  the Galactic disk $\sim 10^8$ years
ago, as is required in Lin's model.

In this paper  we present hydrodynamical  simulations to  analyze the effect
that a single passage of \sgg\ could  have on the  gaseous outer disk of the
Milky Way. By including in the  simulations pressure forces and self-gravity
of the gas, and  adopting the orbit determined  from recent observations, we
model this interaction much more accurately.  More than $\sim 10$ collisions
with   the Galactic disk  are  predicted to  have  occurred over the $\simgt
10$~Gyr  lifetime  of  the  dwarf (IWGIS); however,   due  to the  effect of
dynamical friction during  tidal disruption, one cannot  estimate accurately
the  position and velocity,  or even the  peri-  and apo-Galactic distances,
that \sgg\  had 10~Gyr ago.  We  therefore restrict ourselves in the present
contribution to the consideration  of the effect  of a single encounter with
the Galactic disk;  for  concreteness, we  will study the  aftermath  of the
collision due to occur  $\sim 25$~Myr from  the present time on the opposite
side of the Milky Way from the Sun.

\section{\Sgg: its orbit and mass}

A  comprehensive review  of the   current  observational constraints on  the
nature of \sgr\ is given in IWGIS.  However, for the present study, only its
orbit and its mass are important.

The distance to the \com\ is $25\pm 1 \kpc$ (Ibata  \etal 1994, Mateo \etal\
1995),  while  the  radial  velocity  of  the  \com\   in  a galaxy-centered
non-rotating reference frame is $171 \pm 1 \kms$ (IWGIS).  Several numerical
studies  (Oh \etal\  1995, Piatek    \& Prior 1995,   Johnston \etal\  1995,
Velazquez   \& White 1995)  have shown   that low  mass, initially spherical
stellar  bodies  such as dwarf spheroidal  galaxies  become elongated in the
tidal  field of the  Milky Way,  such that  the elongation aligns  along the
direction   of motion.  This empirical  finding  is only approximately true,
since in the numerical models, the  leading arm of disrupted particles tends
to be nearer to the Galactic center than the lagging arm (see \eg\ Piatek \&
Prior 1995,  their Figure~1).   However, we are  fortunate to  observe \sgg\
from a position which is almost directly opposite the Galactic center to its
present  position, so that its  elongation will in  fact be aligned with its
proper motion vector to very good approximation.  The major axis of \sgg\ is
approximately parallel to  the Galactic coordinate line $l=5^\circ$ (IWGIS),
implying an essentially polar orbit.   Combining these constraints, together
with  a preliminary measurement of the  proper  motion ($2.1 \pm 0.7$~mas/yr
towards the Galactic plane;  Irwin \etal 1996, IWGIS), allows  one to fit an
orbit  for the  dwarf  in   an assumed   Galactic potential.  The   Galactic
potential used to determine the orbit is identical to that detailed below in
Section~3; the center  of mass orbit whose  distance and projected  velocity
best match the observations is presented in Figure~1.

\begin{figure}[htbp]
\psfig{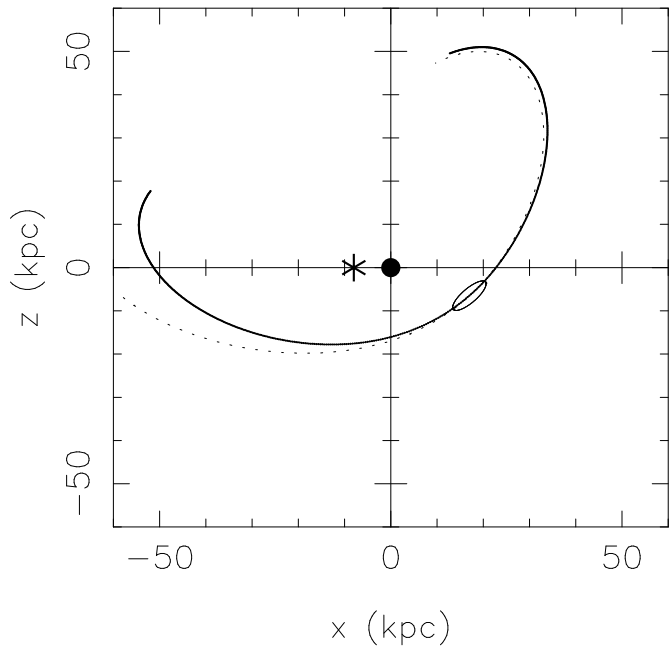}
\caption{The solid line shows the $x$--$z$ plane  of the orbit of a massless
test particle that best fits the kinematic data (see IWGIS for details). The
radial period of this orbit is $0.76 \Gyr$.  The dotted line shows the orbit
taken for the $5 \times 10^9 \msun$ model, where dynamical friction has been
included in the calculations. Both models have been integrated for $1 \Gyr$.
The `star' symbol represents the present position of the Sun, while the open
ellipse (drawn to scale) gives the present position of \sgr.}
\label{FigVibStab} 
\end{figure}

The mass $M_{SG}$ of \sgg\ is a puzzling quantity. One may determine a first
estimate of its  mass by applying the virial  theorem: an object with radial
velocity  dispersion $11.4 \pm 0.7 \kms$  (IWGIS) and a characteristic major
axis diameter $\sim 9.6 \kpc$ (IWGIS) has a  virial mass of $\sim 1.5 \times
10^8  \msun$  (note  that   we have  used the  major   axis length   in this
calculation, which  is likely to give  an overestimate of  the virial mass).
However, as shown  by Velasquez \&  White (1995) and Johnston \etal  (1995),
who  specifically modeled \sgg,  such an extended,  low  mass object is very
fragile,  and would  have been destroyed  by Galactic  tides long before the
present time.  This dilemma may  be solved if  \sgg\ has a substantial  dark
halo, with  mass $\simgt   10^9 \msun$ (IWGIS);  the  low  stellar  velocity
dispersion given this  high mass can  be explained by a radially  increasing
mass to light ratio.

\section{Modeling the interaction}

\subsection{Order-of-magnitude estimates}

We start by  illustrating the  effect of  the passage  of the dwarf   galaxy
through a  thin gaseous disk  using analytical order-of-magnitude arguments.
Neglecting gas dynamics, we can write the momentum transferred to the volume
element of the disk as
\begin{equation}
p \sim \frac{GM_{SG}\rho}{r^2} \cdot \frac{r}{v} \sim \rho u(r),
\end{equation}
where $M_{SG}$ and $v$ are the mass and impact  velocity of the dwarf galaxy
and $r$ is the distance of the disk element from  the place of impact. After
the collision, the gas  element acquires velocity $u(r)  \le v$ which should
bring it to some elevation  $h$ above the disk.  Assuming  the model of  the
Galactic halo potential  described    below, the vertical gradient    of the
potential at the point of impact can be approximated well by the relation
\begin{equation}
|\psi_z| \approx C h,
\end{equation}
where $C \approx 4.4\times  10^{-32}{\rm s}^{-2}$.  The  scale height of the
perturbation caused by the dwarf will therefore be
\begin{equation}
h \sim {1\over {\sqrt{2C}}}{GM_{SG}\over {rv}},
~~~ {\rm for} ~r>R_A,
\end{equation}
where   $R_A=GM_{SG}/v^2$  is the    gravitational   radius  of  Bondi-Hoyle
accretion.  In  our scenario  the orbital  velocity of the  dwarf  is $v=365
\kms$.  With  an extreme case of  $M_{SG}=10^{10}\msun$, this  model gives a
hole $R_A  \approx 300 \pc$ in  radius, while material at  the edge  of this
hole is displaced to  $h \sim  40 \kpc$.   Since $h  \sim r^{-1}$, any  disk
element with impact parameter, say, $r \sim 3\kpc$ can be propelled up to $h
\sim  4  \kpc$ above or  beyond the  disk.  However, this order-of-magnitude
estimate lacks essential physics,  most  importantly dissipation via  shocks
and subsequent radiative cooling.

\subsection{Modeling the \ion{H}{I} disk}

A more careful treatment is required to predict quantitatively the effect of
the collision on the gaseous disk.  Walker \etal\ (1996) presented an N-body
simulation to study the interaction  of a disk--halo--satellite system in  a
merger which does   not destroy the disk.  After   such an event,  the  disk
appears   to be  of  an earlier  Hubble  type than  its  progenitor,  with a
substantial amount of heating having been pumped into the disk.  This energy
will clearly  alter the physical  state of the gas in  the disk resulting in
more physical diffusion than in purely N-body  simulations.  To address this
issue, we modeled the   \ion{H}{I} as a continuous  self-gravitating  fluid,
using a smooth particle hydrodynamics  (SPH) tree-code kindly provided by M.
Steinmetz (Steinmetz \& M\"uller 1993).

For all of our models of the \ion{H}{I} disk, we assume an initially uniform
temperature $T_{gas} \sim 100 {\rm K}$.  We take an exponentially decreasing
surface  mass  density beyond $R=10\kpc$  with   scale length  $6 \kpc$ (our
parameterization  of  the data  of  Burton  \&  te Lintel Hekkert   1986), a
Gaussian density distribution in the direction perpendicular to the Galactic
plane with a scale height of $500\pc$ (the  measured value at $R=10\kpc$ ---
Burton (1992)), and a  number density of  \ion{H}{I} at $z=0$ of  $1.0 ~{\rm
atoms~cm^{-3}}$ (so   as to yield  the surface  mass  density  at $R=10\kpc$
measured by Burton  \& te Lintel  Hekkert 1986).  We chose  to set the outer
limit of the disk  at $R=30\kpc$. We also chose,   for tractability, not  to
populate the \ion{H}{I} disk inside $R=10\kpc$,  to avoid having to simulate
the complex  interaction of  the gaseous  disk with the  stellar disk.   The
total mass in \ion{H}{I} in the model is then $8.5 \times 10^9 \msun$.

Artificial viscosity  in the simulation  is corrected for shearing, as given
by Benz (1990),  which would otherwise result  in  a rapid transfer of  mass
from the outer to the inner regions of the disk.

One  has to make sure that  the surface number density  of  particles in the
disk is high enough to resolve the impact area.  Our criterion is to have at
least $100$ particles in the disk impact region corresponding to the size of
the  dwarf.  We therefore  chose to  populate  the annulus representing the
\ion{H}{I}  gas   with  $3\times  10^4$  particles   whose  masses  decrease
exponentially from  $R=10 \kpc$ with  a scale length of   $6 \kpc$, giving a
uniform surface number density.

\subsection{Cooling of the atomic hydrogen}

An appropriate thermal model  for the atomic hydrogen  is required, so as to
mimic the existing temperature and density distributions of the gaseous disk
as closely as  possible.  In nature, the temperature  of the gas follows the
equilibrium  temperature   set by  the  interplay  of   heating and  cooling
processes.   However, this process is  very complex: the  cooling rate for a
given temperature, density and  chemical composition  depends on the  number
density of free  electrons, which in turn,  depends on the flux  of ionizing
radiation  (from cosmic rays, star  formation, supernovae, and so on), which
is poorly constrained   in  the outer regions   of  the Galaxy.   The   most
problematic  issue is  that  the ISM  is likely   to be  optically  thick to
radiation  in  some important  emission  lines;  developing  a cooling  rate
algorithm with the necessary  radiative transfer calculations is far  beyond
the scope of the present contribution.

To  make the problem tractable, we  instead assume an  ISM equilibrium state
given  by  Scheffler  \& Els\"asser  (1988),  which  is  based on  the local
chemical composition  and heating  rate   in the absence  of  hydrodynamical
heating.  The settling of the gas onto this equilibrium curve will depend on
the amount of dissipational heating, which in our case, is  given by the SPH
algorithm.  The   cooling  timescale is then just   taken  from the explicit
cooling functions for atomic hydrogen of Scholz \&  Walters (1991), with the
modification  that   the    cooling  time  may     not   be   longer    than
$T_{cool}=10^5$~years. It was found that  the results were not sensitive  to
the choice of this upper bound in the range  $10^5 < T_{cool}< 10^6$, in the
sense that the configuration of the disk at the end point of the simulations
were qualitatively indistinguishable.

\subsection{The Galactic potential}

Ideally, one would prefer to model the  Galaxy with a ``live'' halo, stellar
disk, bulge and spheroid.  However, the computational resources required for
such a   treatment are beyond our  present  capabilities.  Instead,  the SPH
scheme was altered to include  a fixed  potential for the   Milky Way and  a
moving  potential for the  dwarf galaxy.  The potential of  the Milky Way is
derived from the mass model  of Evans \& Jijina  (1994).  In this model, the
disk component,  described by  a double exponential   disk, has radial scale
length   $h_R=3.5 \kpc$   and  a  Solar  neighborhood  surface   density  of
$\Sigma_0=48 \msun/\pc^2$. We further assume  that the vertical scale length
of the disk is $h_z=0.25 \kpc$, and that the density falls to zero at $R = 5
h_R$.  The potential corresponding to  this density distribution is found by
multipole expansion of the Poisson equation using  an algorithm described in
Englmaier (1997).  The halo component is described by a `power-law' halo, so
the potential has the following analytical expression:
$$
\Psi={ {v_0^2 R_c^\beta / \beta} \over 
{ ( R_c^2 + R^2 + z^2 q^{-2} )^{\beta/2} } }, {\hskip 1cm} \beta \ne 0,
$$
where $R$ and   $z$  are Galactocentric  cylindrical  coordinates,  the core
radius $R_c = 2\kpc$, $v_0 = 138 \kms$, the exponent $\beta = -0.2$, and the
oblateness parameter $q = 1$.

A shortcoming of this approach is that the fixed halo behaves differently to
a ``live'' halo that is free to respond to the passage of the massive dwarf.
A ``live'' halo would respond (in a way that is dependent on the density and
kinematics  of the halo, and  the mass and  velocity  of the  object) to the
extra gravitational attraction, leaving a  wake of halo material behind  the
object. This   wake would also  interact  with the  \ion{H}{I} disk.  So, in
reality,  the  interaction of \sgg\  with  the Galactic disk  should be more
violent than found here using a fixed halo model.

\subsection{Modeling the dwarf galaxy}

To  obtain a  good representation of   the dwarf  galaxy,  one would ideally
construct a set  of spherical stellar  models,  evolve them in the  Galactic
potential for $\sim  10$~Gyr, and choose the model  that best reproduces the
observations.  However, as yet  no N-body model  has been found that is able
to survive Galactic tides and  also give a  reasonable approximation to  the
observed distance, radial velocity dispersion  and radial velocity gradient.
In the absence of such a self-consistent N-body  model, we account for \sgg\
by including a Plummer sphere potential $\Psi_{SG}=- G M_{SG} / \sqrt( r^2 +
r_0^2 )$, which progresses along a predetermined orbit. Here $r$ is a radial
distance from the guiding center of the orbit, and $r_0 = 1 \kpc$. Note that
spherical models are not a good representation of the present shape of \sgg,
since it is significantly elongated (axis  ratios 3:1:1; IWGIS), however the
difference to the perturbation on the Galactic gas is likely to be small.

For  a chosen model mass   $M_{SG}$, we calculate  the  orbit  in the  above
Galactic potential  whose projected velocity best   fits the kinematic data;
dynamical friction  is taken into  account using the Chandrasekhar dynamical
friction formula (see \eg\ Binney \& Tremaine 1987).  So  as to minimize the
perturbations  on the  Galactic  disk   in the  initial  evolution  of   the
simulation, at the beginning of the integration, the dwarf galaxy models are
placed at the  position where the \com\ of  the model was $4.4 \times  10^8$
years ago.

\section{Simulation results}

We first followed  the evolution of the  gaseous disk in  the absence of the
dwarf.  Over the course of a $5 \Gyr$ integration,  the disk was observed to
develop small-scale spiral   features (high $m$-mode instabilities),  but it
remained axisymmetric,  flat (no simulation  particle rose beyond  $|z|= 1.5
\kpc$), and in particular, it was devoid of  large spiral arms (low $m$-mode
features).

Two other simulations --- all integrated for $1.75  \Gyr$ --- were performed
with  dwarf  galaxy  masses $M_{SG}=10^9$ and  $5   \times 10^9 \msun$.   As
discussed above,  \sgg\ should have a  mass of $M_{SG}  \sim  10^9 \msun$ to
have survived in  the Galaxy's tidal  field  over its  lifetime of $\sim  10
\Gyr$.  Taking into account that a live Galactic  halo will only augment any
perturbations  to the HI disk,  $M_{SG}=5 \times 10^9  \msun$ is a reasonable
upper  limit  to be considered.  It   seems plausible that  a more realistic
simulation with a  time-dependent, inhomogeneous halo  able to respond to the
passage of a  $\sim 2-3 \times  10^9 \msun$ dwarf  galaxy, would demonstrate
results similar to our fixed-halo $M_{SG}=5 \times 10^9 \msun$ simulation.

\begin{figure*}[htb]
\psfig{figure=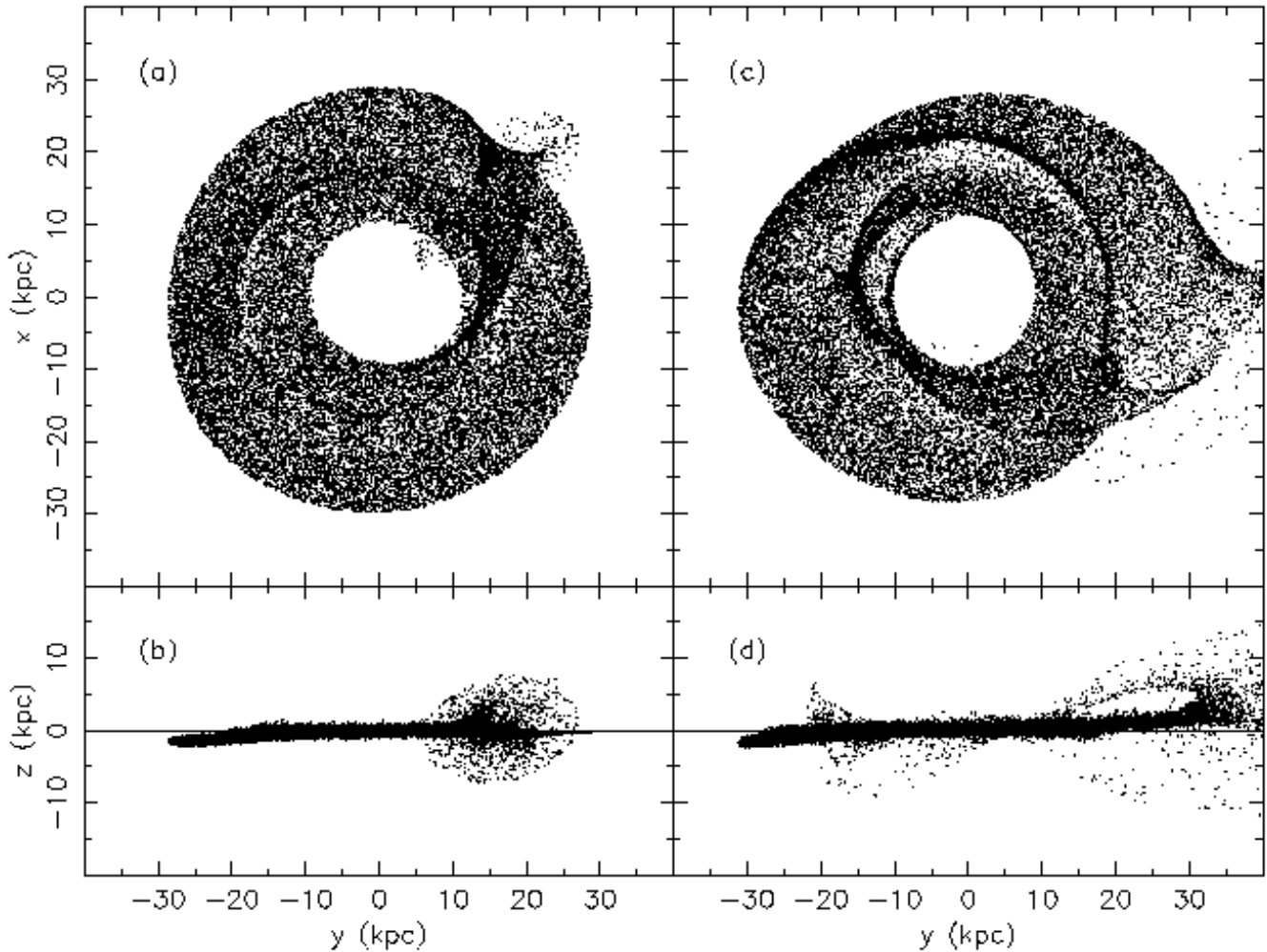,width=18.0cm,angle=270}
\caption[ ]{The structure of the \ion{H}{I} disk is  shown for the $M_{SG}=5
\times 10^9 \msun$ simulation, at simulation times of $0.5 \Gyr$ (just after
the collision) and $1 \Gyr$ in the left- and right-hand panels respectively.
The coordinate system is inertial, chosen such that the Galactic center lies
at the origin and the {\it present} position of the Sun is at (-8,0,0).  The
Galactic plane is  viewed from above, so the  sense of  Galactic rotation is
clockwise  in the upper  panels.  The point  of impact  with  the disk is at
$x=22.5\kpc$, $y=0\kpc$.}
\end{figure*}

We  now focus on the results  of the  simulation  with $M_{SG}=5 \times 10^9
\msun$. Just  before the first  collision with  the dwarf, the  disk remains
smooth, and  only  slightly perturbed (the  mean level  of the outer gaseous
disk droops towards  the dwarf by $\sim 300\pc$).   In Figures~2a and 2b, we
display the   $x$--$y$  and $y$--$z$ plane structure    of  the Milky  Way's
\ion{H}{I} disk just after the encounter at $T=0.5 \Gyr$ (\ie\ $60$~Myr from
the present time).  The  collision  produces a  plume of ejected  particles,
seen clearly in   Figure~2b.  The disk  becomes  deformed near the point  of
collision,  with  vertical   displacement  that follows    approximately the
$r^{-1}$ relation predicted by  the order of magnitude calculation presented
above.  Furthermore, on  the  opposite side of  the  Galaxy to the  point of
impact, the the  \ion{H}{I} layer  has been pulled   down from the  Galactic
plane by $\simgt 1 \kpc$ due to the attractive influence of the dwarf.

The structure of  the  disk after $1\Gyr$  is shown  in Figures~2c  and  2d.
Figure~3  shows   the corresponding  column   density  of  \ion{H}{I}  in  a
Galactocentric annulus  of  width $1  \kpc$  centered  on the  point  of the
collision.  At that  time,  $0.5 \Gyr$   after  the collision,  the  disk is
clearly non-axisymmetric: large  spiral--like arms dominate the morphology of
the  outer  regions, while  the whole  disk  appears lopsided.  The vertical
displacement of  the gas resembles   the observed warp,  with main  features
separated by $\sim 180$ degrees.

\begin{figure*}[htb]
\psfig{figure=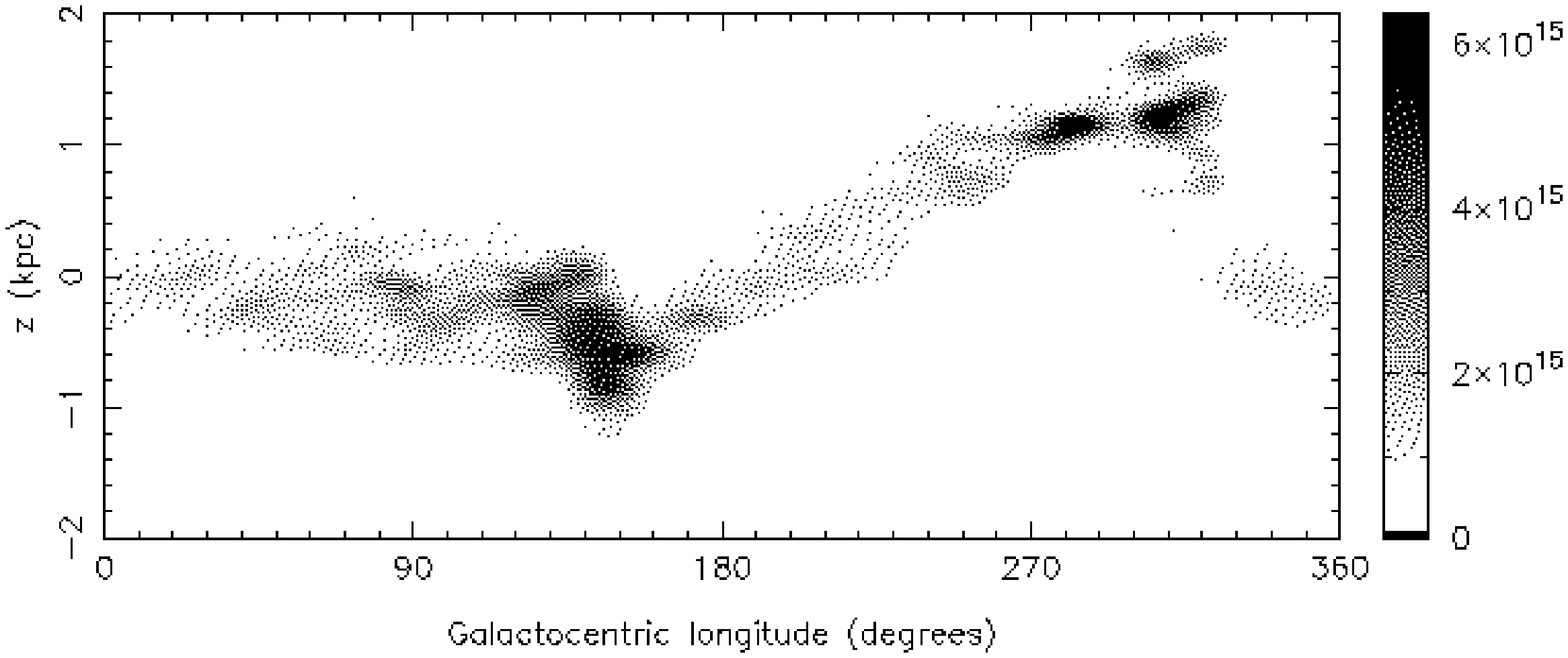,width=18.0cm,angle=270}
\caption[  ]{The column density map   (in ${\rm atoms/cm^2}$) of  \ion{H}{I}
particles in a cylinder with $1 \kpc$  thick walls centered at $R=22.5 \kpc$
(the  point of  impact), as seen   from the Galactic   center, is displayed.
Here, the dwarf galaxy model has mass  $M_{SG}=5 \times 10^9 \msun$, and the
simulation time is $1 \Gyr$.  The angle termed 'Galactocentric longitude' is
measured anti-clockwise  in the sense  of the upper  panels of Figure~2; the
direction   corresponding  to the  {\it  present}  position  of  the  Sun is
$0^\circ$.}
\end{figure*}

The perturbation on the \ion{H}{I} disk, it transpires, is a strong function
of  mass --- the effect of  the $M_{SG}=10^9 \msun$  on the structure of the
disk being just noticeable, while the model with $M_{SG}=5\times 10^9 \msun$
produces a noticeable warp-like structure in the  disk. The degree of warping
depends on the  amount of physical and numerical  diffusion in the model. In
Figure~4 we demonstrate  that the adopted equation  of state (ideal gas with
cooling to  an  equilibrium state) together  with the   corrected artificial
viscosity, in effect, leads to very little numerical diffusion allowing most
of the gas in the plane of the  disk to sit at  temperatures of $\approx 100
{\rm K}$. There is substantial  heating only in disk  shocks and tidal tails
which follow the dwarf galaxy into the halo, but then this gas cools down to
produce two atomic components of the ISM (diffuse and cold phases).

\begin{figure}[htb]
\psfig{figure=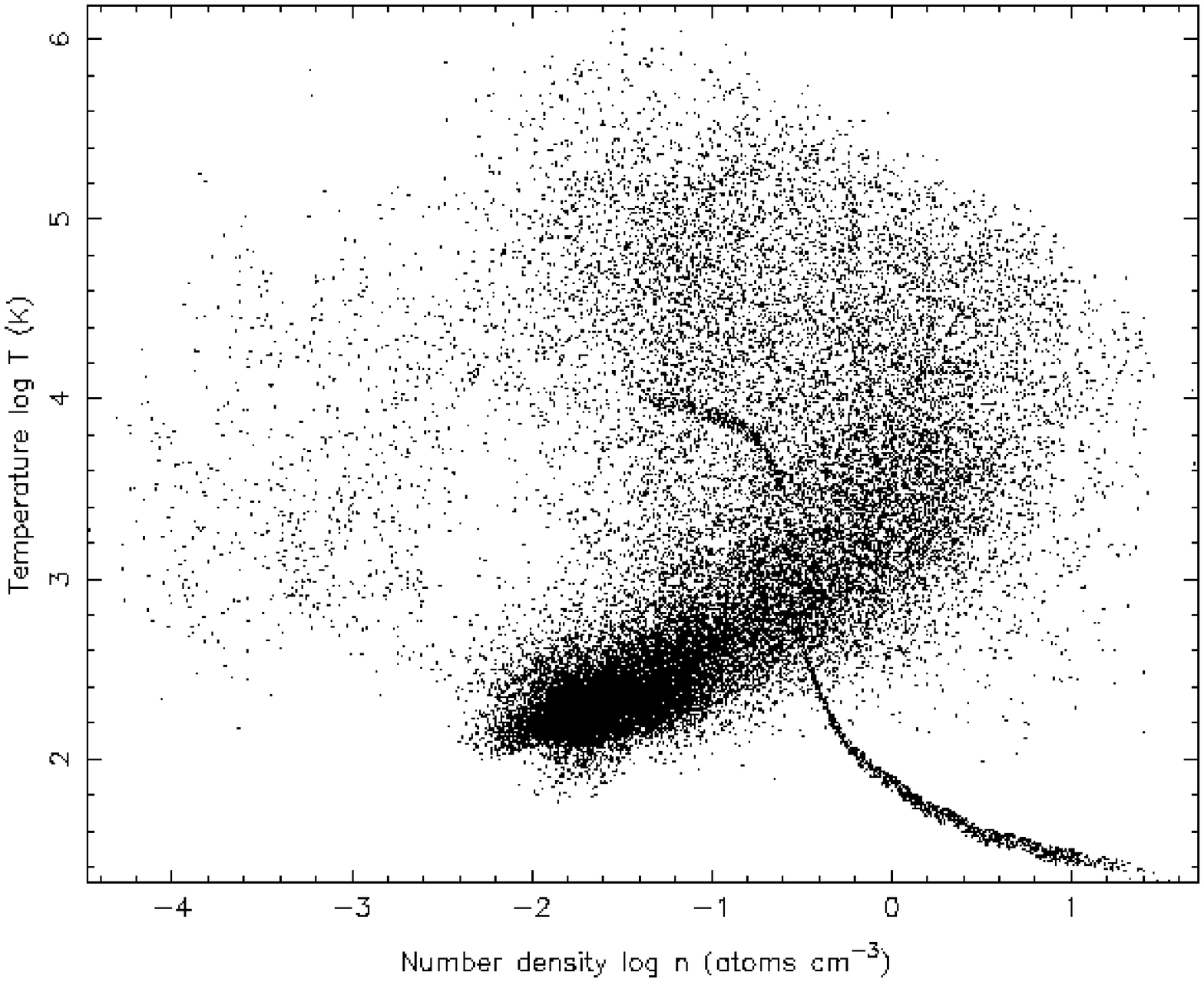,width=8.8cm,angle=270}
\caption[  ]{The  number density--temperature  diagram corresponding  to the
simulation   displayed  in  Figures~2a  and  2b.   At  any  time  during the
simulations, the majority of particles stay at low temperatures in the lower
left corner  of the diagram.   A  minor  portion of  the  material is  being
constantly heated through dissipational processes.  The heated gas is forced
to cool down (see details in the text) to the  equilibrium curve which is in
turn separated into a cold ($\log T \sim 1.8$) and warm  ($\log T \sim 3.8$)
component, as expected.  Ideally, one would like to mimic the density map of
the ISM, comparing   models to the  observed mass  distribution inside dense
clouds and  in the form of diffuse  gas.  (Note that the particles displayed
in the plot have variable masses.)}
\end{figure}

The perturbations   induced  by the dwarf on    the Galactic \ion{H}{I}  are
long-lived, as we show in Figure~5,  which displays the configuration of the
SPH particles at the end of the $M_{SG}=5\times 10^9 \msun$ simulation, just
before the third impact  with the Galactic  disk.  As discussed above, \sgg\
has likely had many collisions with the Galactic disk in  the past, and will
soon crash through the disk yet again.   Thus, the outer \ion{H}{I} disk may
presently  be quite disturbed, if the  mass of \sgg\  is indeed in the range
considered here.  Once the mass of the dwarf  is better constrained, further
simulations,  with  a live halo,  should   be performed   to  obtain a  more
realistic estimate of the magnitude of these disturbances.

\begin{figure*}[htb]
\psfig{figure=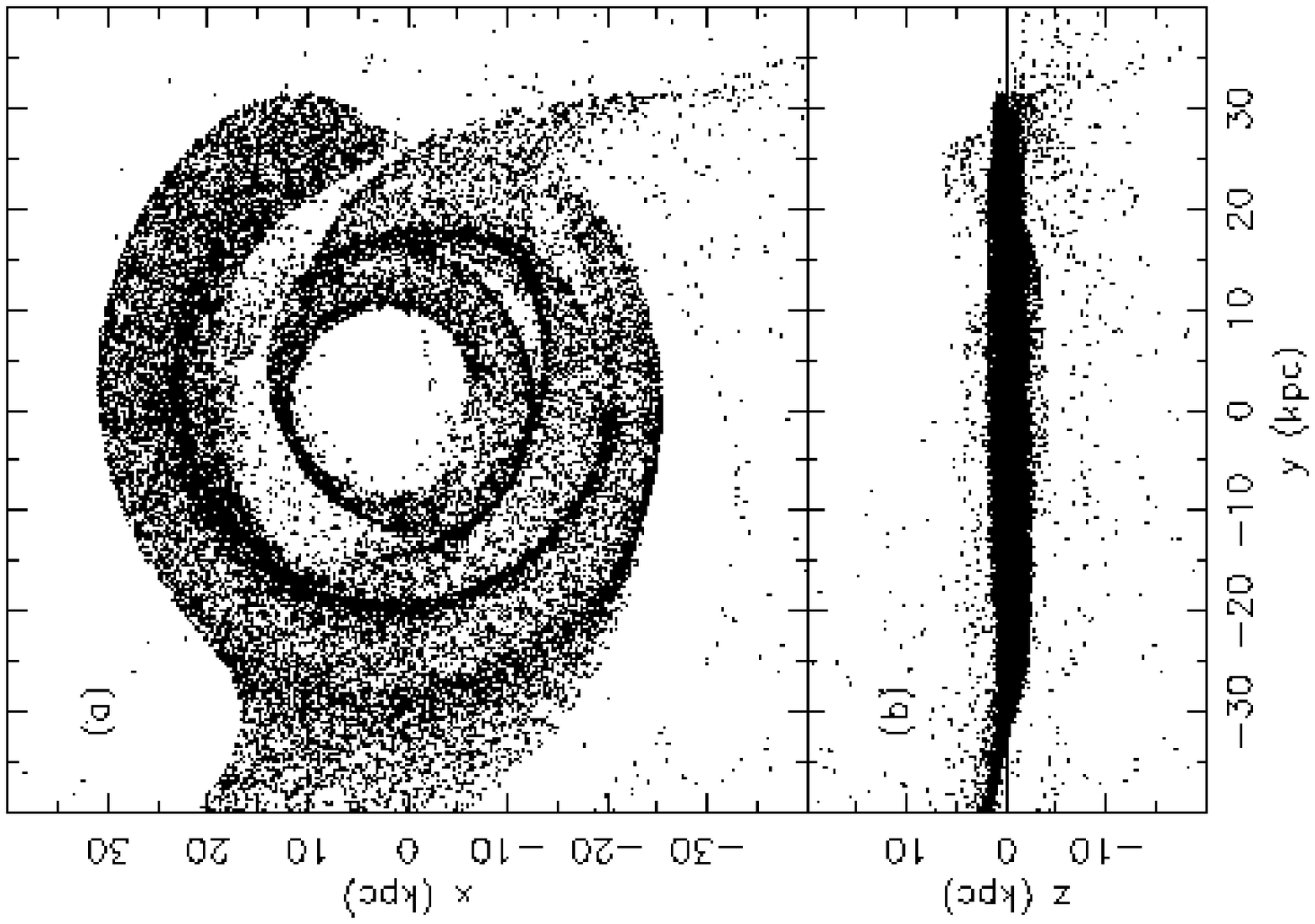,width=8.8cm,angle=270}
\caption[ ]{As Figure~2, but at the simulation time $1.75 \Gyr$, just before
a third collision  of the dwarf with  the  Galaxy. The disturbances  clearly
remain prominent for longer than the period between collisions.}
\end{figure*}

\section{Conclusions}

Many mechanisms  have  been proposed   to explain the  --- very   common ---
phenomenon of warps  in gaseous  disks (see  Binney 1992),  with perhaps the
most   reasonable   explanation being  that   warps are   simply the natural
oscillatory response of the gaseous disk to small perturbations. As such, it
is conceivable that \sgg\ may provide a perturbing  influence which helps to
excite periodically the Galactic warp, especially  since it seems that warps
are transient  phenomena and need to  be  maintained (Binney 1992).

Perhaps most importantly,  our  simulations show  that  it  is necessary  to
include   the perturbative  effect of \sgg\    to fully understand  the star
formation history and  spiral structure  of the  outer Galactic  disk.  In a
subsequent contribution,   we will improve on  the  accuracy  of the present
simulations by including  a live Galactic  halo; without accounting for  the
effects  of a  real  halo, it is probably  premature  to attempt  a detailed
comparison of the models to the observed \ion{H}{I} distribution.

An interesting prospect is that it may be  possible to detect indirectly the
presence of  dwarf  galaxies from their  effect on  the \ion{H}{I} disks  of
their giant neighbors. For instance,  a relatively massive galaxy similar to
\sgg\ would  be extremely  hard  to detect  (due to  the  very  low  surface
brightness)  if it were  orbiting an  external galaxy  at a  distance beyond
which its brightest giants could  be resolved, whereas  the effect caused on
the structure of  its neighbor's  gaseous disk  could be readily  observable
(given   a  sufficiently  high mass).   Also,   recent numerical simulations
(Johnston   \etal\  1996) suggest that matter    disrupted from former dwarf
galaxies can  form very  long-lived streams in   the halo of   their massive
companions.  \ion{H}{I} disks  could also be  affected  by these structures,
depending on their  mass density and clumpiness.  This  may yield a means to
quantify  the  present merging  rate (see,  e.g.,  Zaritsky \&  Rix 1997), a
parameter of great value to galaxy formation theory.

\bigskip

\begin{acknowledgements}

We are very grateful to Matthias Steinmetz for kindly allowing us the use of
his {\tt SPHINX} code,  and  to J.~Auman, G.~Fahlman, G.~Lewis,  T.K.~Menon,
H.~Richer and D.~Scott for helpful  discussions.  RAI expresses gratitude to
the   Killam  Foundation (Canada)  and   to the   Fullam  Award for support.
\vfill\eject

\end{acknowledgements}

\end{document}